\documentstyle[12pt]{article}
\newcommand{\noi}{\noindent}
\newcommand{\h}{\hspace*{4ex}}

\newcommand{\cent}{\centerline}
\newcommand{\vs}{\vspace*}

\begin{document}

\baselineskip 0.6cm

\begin{center}

{\large {\bf The scientific manuscripts left unpublished by Ettore
Majorana}
 (with outlines of his life and work)}$^{\: (\dag)}$
\footnotetext{$^{\: (\dag)}$  E-mail addresses for contacts: sesposit@na.infn.it [SE];
 \ recami@mi.infn.it [ER]}

\end{center}

\vs{3mm}

\cent{ Erasmo Recami }

\vs{1mm}

\cent{{\em Facolt\`a di Ingegneria, Universit\`a statale di Bergamo,
Bergamo, Italy;}}
\cent{{\rm and} {\em INFN--Sezione di Milano, Milan, Italy.}}

\vs{1mm}

\centerline{\rm and}

\cent{ Salvatore Esposito }

\vs{1mm}

\cent{{\em Universit\`a di Napoli, Neaples, Italy;}}
\cent{{\rm and} {\em INFN--Sezione di Napoli, Neaples, Italy.}}

\vs{5mm}

{{\bf Abstract  \ --} \ We present here a panoramic view of the
main {\em scientific manuscripts} left unpublished by the
brightest Italian theoretical physicist of the XX century, Ettore
Majorana. \ We deal in particular: \ (i) with his very original
``study" notes (the so-called {\em Volumetti}), already published
by us in English, in 2003, c/o Kluwer Acad. Press, Dordrecht \&
Boston, and in the original Italian language, in 2006, c/o
Zanichelli pub., Bologna, Italy; \ and  \ (ii) with a selection of
his research notes (the so-called {\em Quaderni}), that we are
going to publish c/o Springer, Berlin. \  We seize the present
opportunity for setting forth also some suitable ---scarcely
known--- information about Majorana's life and work, on the basis
of documents (letters, testimonies, various documents...)
discovered or collected by ourselves during the last decades. \ A
finished, enlarged version of this paper will appear, as a Preface
by the editors, at the beginning of the coming book {\em Ettore
Majorana -- Unpublished Research Notes on Theoretical Physics},
edited by S.Esposito, E.Recami, A.van der Merwe and R.Battiston,
to be printed by Springer verlag. [Let us recall that almost all
the {\em biographical} documents regarding Ettore Majorana, photos
included, are protected by copyright in favor of Maria Majorana
together with the present authors (mainly ER), and partly with the
publisher Di Renzo (www.direnzo.it), and cannot be further
reproduced without the written permission  of the right holders].

\newpage

\section*{1. Ettore Majorana's life and work}

\

\h     ``Without listing his works, all of which are highly
notable both for their originality of the methods utilized as well
as for the importance of the achieved results, we limit ourselves
to the following:

\h     ``In modern nuclear theories, the contribution made by this
researcher to the introduction of the forces called `Majorana
forces' is universally recognized as the one, among the most
fundamental, that permits us to theoretically comprehend the
reasons for nuclear stability.  The work of Majorana today serves
as a basis for the most important research in this field.

\h     ``In atomic physics, the merit of having resolved some of
the most intricate questions on the structure of spectra through
simple and elegant considerations of symmetry is due to Majorana.

\h     ``Lastly, he devised a brilliant method that permits us to
treat the positive and negative electron in a symmetrical way,
finally eliminating the necessity to rely on the extremely
artificial and unsatisfactory hypothesis of an infinitely large
electrical charge diffused in space, a question that had been
tackled in vain by many other scholars.'' [4]

\

\noi With this justification, the judging committee of the 1937
competition for a new full professorship in theoretical physics at
Palermo, chaired by Enrico Fermi (and including Enrico Persico,
Giovanni Polvani and Antonio Carrelli) suggested the Italian
Minister of National Education to appoint Majorana,
``independently of the competition rules, as full-professor of
theoretical physics in a university of the Italian
kingdom\footnote{Which happened to be that of Naples.} because of
his high and well-deserved reputation'' [4]. Evidently, to get
such a high reputation there were enough the {\em few} papers that
the Italian scientist had chosen to publish.  It is interesting to
note that proper light was shed by Fermi on Majorana's symmetrical
approach to electrons and anti-electrons (today climaxing in its
application to neutrinos and anti-neutrinos) and on its ability in
eliminating the hypothesis known as the ``Dirac sea'':  An
hypothesis that Fermi defined as ``extremely artificial and
unsatisfactory,'' despite the fact that in general it had been
uncritically accepted. Though, one of the most important works of
Ettore Majorana, the one that introduced his ``infinite-components
equation,'' was not mentioned: it had not been yet understood,
even by Fermi and colleagues.

\h    Bruno Pontecorvo, a younger colleague of Majorana at the Institute of Physics in
Rome, in a similar way recalled that [2]: ``some time after his entry into Fermi's group,
Majorana already possessed such an erudition and had reached such a high level of
comprehension of physics that he was able to speak on the same level with Fermi about
scientific problems.  Fermi himself held him to be the greatest theoretical physicist of
our time.  He often was astounded [...].''

\

\h Ettore Majorana's fame solidly rests on testimonies like these, and even
more on the following ones.

\h At the request of Edoardo Amaldi [1], Giuseppe Cocconi wrote
from CERN (July 18, 1965):

\

\h     ``In January 1938, after having just graduated, I was
invited, essentially by you, to come to the Institute of Physics
at the University in Rome for six months as a teaching assistant,
and once I was there I would have the good fortune of joining
Fermi, Gilberto Bernardini (who had been given a chair at Camerino
university a few months earlier) and Mario Ageno (he, too, a new
graduate), in the research of the products of disintegration of
$\mu$ ``mesons'' (at that time called mesotrons or yukons), which
are produced by cosmic rays. [...]

\h  ``A few months later, while I was still with Fermi in our
workshop, news arrived of Ettore Majorana's disappearance in
Naples.  I remember that Fermi busied himself with telephoning
around until, after some days, he had the impression that Ettore
would never be found.

\h  ``It was then that Fermi, trying to make me understand the
significance of this loss, expressed himself in quite a peculiar
way; he who was so objectively harsh when judging people.  And so,
at this point, I would like to repeat his words, just as I can
still hear them ringing in my memory: `Because, you see, in the
world there are various categories of scientists: people of a
secondary or tertiary standing, who do their best but do not go
very far.  There are also those of high standing, who come to
discoveries of great importance, fundamental for the development
of science' (and here I had the impression that he placed himself
in that category).  `But then there are geniuses like Galileo and
Newton.  Well, Ettore was one of them.  Majorana had what no one
else in the world had [...]'.''

\

\noi Enrico Fermi, who was rather severe in his judgements, again
expressed himself in an unusual way on another occasion.  On July
27, 1938 (after Majorana's disappearance, which took place on
March 26, 1938), writing from Rome to the Prime Minister Mussolini
in order to ask for an intensification of the search for Ettore,
he stated:

\

\h ``I do not hesitate to declare, and it would not be an
overstatement in doing so, that of all the Italian and foreign
scholars that I have had the chance to meet, Majorana, for his
depth of intellect, has struck me the most.'' [4]

\

\noi But, nowadays, a part of the interested scholars may find difficult to catch
Majorana's ingeniousness by basing themselves only on his few published papers (listed
below), most of them originally written in Italian and not easy to be traced; only three
articles of his having been translated into English [26,9,10,11,12] in the past.
Actually, only in 2006 the Italian Physical Society did eventually publish a book with
the Italian and English versions of Majorana's articles [13].

\h Anyway, the Italian scientist has left also a lot of
unpublished manuscripts about his studies and research, mainly
deposited at the Domus Galilaeana of Pisa (in Italy), which help
illuminating us on his abilities as a theoretical physicist, and
mathematician too.

\h The year 2006 has been the 100th anniversary of the birth of
Ettore Majorana: probably the brightest Italian theoretician of
the XX century, even if to some people Majorana is known mainly
for his mysterious disappearance, in 1938, when he was 31. \ For
celebrating such a centenary, we had been working ---among the
others--- about selection, study, typographical setting in
electronic form, and translation into English of the most
important research notes left unpublished by Ettore Majorana: His
so-called {\em ``Quaderni"} (booklets); leaving aside, for the
moment, the noticeable set of loose sheets that constitute a
conspicuous part of Majorana's manuscripts. \ Such a selection is
going to be published for the first time, with some understandable
delay, in a coming book. \ In a previous volume [15], that
hereafter we shall call {\em ``Volume I"}, we have analogously
published for the first time the material contained in {\em
different} Majorana's booklets ---the so-called ``Volumetti''---,
which had been written down by him mainly while {\em studying}
physics and mathematics as an Enrico Fermi's student and
collaborator. \ Even if ``Volume I" did already contain many
highly original exploits, the preparation of the present book
remained nevertheless as a rather necessary enterprize, since the
research-notes appearing below are even more (and often
exceptionally) interesting, showing at a larger degree Majorana's
genius.  Many of the following results are still important, and a
novelty, for contemporary theoretical physics, even after about
eighty years.

\

\section*{2. A further historical prelude}

\noi For non specialists, the name of Ettore Majorana is
frequently associated with his mysterious disappearance from
Naples, on March 26, 1938, when he was only 31; afterwards, in
fact, he was never seen again.

\h But the myth of his ``disappearance'' [4] has contributed to nothing but the fame he
was entitled to, for being a genius well ahead of his time.

\h Ettore Majorana was born on 5 August 1906 at Catania, Sicily
(Italy), to Fabio Majorana and Dorina Corso. The fourth of five
sons, he had a rich scientific, technological and political
heritage: three of his uncles had become vice-chancellors of the
University of Catania and members of the Italian parliament, while
another, Quirino Majorana, was a renowned experimental physicist,
who had been, by the way, a former president of the Italian
Physical Society.

\h Ettore's father, Fabio,  was an engineer who had founded
the first telephone company in Sicily and who went on to become chief
inspector of the Ministry of Communications. \  Fabio Majorana was
responsible for the education of his son in the first years of his
school-life, but afterwards Ettore was sent to study
in a boarding school at Rome. Eventually, in 1921, the whole family moved
from Catania to Rome. Ettore finished his high-school in 1923 when he was
17, and then joined the Faculty of Engineering of the local university, where
he excelled, and counted Giovanni Gentile Jr., Enrico Volterra, Giovanni
Enriques, and future Nobel laureate Emilio Segr\`e among his friends.

\h In the spring of 1927 Orso Mario Corbino, the director of the
Institute of Physics at Rome and an influential politician (who
had succeeded in elevating to full-professorship the 25 years old
Enrico Fermi, just with the intention to allow the Italian physics
to make a quality jump) launched an appeal to the students of
Engineering, inviting the most brilliant young minds to study
physics. \ Segr\`e and Edoardo Amaldi rose to the challenge,
joining Fermi and Franco Rasetti's group, and telling them of
Ettore's exceptional gifts. After some encouragement from Segr\`e
and Amaldi, Majorana eventually decided to meet Fermi in the
autumn of that year.

\h     The details of Majorana and Fermi's first meeting were
narrated by E. Segr\'e [3], Rasetti, and Amaldi. The first
important work written by Fermi in Rome, on the statistical
properties of the atom, is today known as the Thomas-Fermi method.
Fermi had found that he needed the solution to a non-linear
differential equation characterized by unusual boundary
conditions, and in a week of assiduous work he had calculated the
solution with a little hand calculator.  When Majorana met Fermi
for the first time, the latter spoke about his equation, and
showed his numerical results.  Majorana, who was always very
skeptical, believed Fermi's numerical solution to be probably
wrong. He went home, and solved Fermi's original equation in
analytic form, evaluating afterwards the solution's values without
the aid of any calculator. Next morning he returned to the
Institute and skeptically compared the little piece of paper on
which he had written his results to Fermi's notebook, and found
that their results coincided exactly, he could not hide his
amazement; and decided to go on from Engineering to Physics. \ We
have indulged in the foregoing anecdote since the pages on which
Majorana solved Fermi's differential equation were found by one of
us (S.E.) years ago. And recently [22] it has been explicitly
shown that he followed that night two independent paths, the first
of them leading to an Abel equation; and the second one devising a
method still unknown to mathematics. More precisely, Majorana
arrived at a series solution of the Thomas-Fermi equation by using
an original method that applies to an entire class of mathematical
problems.  More in general, while some of Majorana's results
anticipated by several years those of renowned mathematicians or
physicists, several others (including his final solution to the
mentioned equation) have not been obtained by anyone else since:
Such facts are further evidence of Majorana's brilliance.

\

\section*{3. Majorana's published articles}

\noi Majorana published few scientific articles: nine, actually,
besides his sociology paper entitled ``Il valore delle leggi
statistiche nella fisica e nelle scienze sociali'' (The value of
statistical laws in physics and the social sciences), which was
however published not by Majorana but (posthumously) by G. Gentile
Jr., in {\it Scientia} [{\bf 36} (1942) 55-56], and much later
translated into English.
%
\ Majorana switched from the Engineering to the Physics studies in 1928 (the year in
which he published his first article, written in collaboration with his friend Gentile)
and then went on to publish his works in theoretical physics only for few years,
practically only until 1933. \ Nevertheless, even his {\em published} works are a mine of
ideas and techniques of theoretical physics that still remains largely unexplored. \ Let
us list his nine published articles, which only in 2006 have been eventually reprinted
together with their English translation [13]:

\

\begin{enumerate}
\item[{(1)}]
  ``Sullo sdoppiamento dei termini Roentgen ottici a causa
dell'elet\-tro\-ne rotante e sulla intensit\`a delle righe del
Cesio,'' in collaboration with Giovanni Gentile Jr., {\em
Rendiconti Accademia Lincei} {\bf 8} (1928) 229-233.

\item[{(2)}] ``Sulla formazione dello ione molecolare di He,''
{\em Nuovo Cimento} {\bf 8} (1931) 22-28.

\item[{(3)}] ``I presunti termini anomali dell'Elio,'' {\em Nuovo
Cimento} {\bf 8} (1931) 78-83.

\item[{(4)}] ``Reazione pseudopolare fra atomi di Idrogeno,'' {\em
Rendiconti Accademia Lincei} {\bf 13} (1931) 58-61.

\item[{(5)}] ``Teoria dei tripletti {\em P'} incompleti,'' {\em
Nuovo Cimento} {\bf 8} (1931) 107-113.

\item[{(6)}] ``Atomi orientati in campo magnetico variabile,''
{\em Nuovo Cimento} {\bf 9} (1932) 43-50.

\item[{(7)}] ``Teoria relativistica di particelle con momento
intrinseco arbitrario,'' {\em Nuovo Cimento} {\bf 9} (1932)
335-344.

\item[{(8)}] ``\"Uber die Kerntheorie,'' {\em Zeitschrift f\"ur
Physik } {\bf 82} (1933) 137-145; ``Sulla teoria dei nuclei,''
{\em La Ricerca Scientifica} {\bf 4}(1) (1933) 559-565.

\item[{(9)}] ``Teoria simmetrica dell'elettrone e del positrone,''
{\em Nuovo Cimento} {\bf 14} (1937) 171-184.
\end{enumerate}

\

\h While still an undergraduate, then, in 1928 Majorana published
his first paper, (1), in which he calculated the splitting of
certain spectroscopic terms in gadolinium, uranium and caesium,
due to the spin of the electrons. At the end of that same year,
Fermi invited Majorana to give a talk at the Italian Physical
Society on some applications of the Thomas-Fermi model [23]
(attention to which has been called by F. Guerra and N. Robotti) .
Then on 6 July 1929, Majorana got his master degree in physics,
with a dissertation having as a subject ``The quantum theory of
radioactive nuclei.''

\h By the end of 1931 the 25-year-old physicist had published two
articles, (2) and (4), on the chemical bond of molecules, and two
more papers, (3) and (5) on spectroscopy, one of which, (3),
anticipated results later obtained by a Samuel Goudsmith's
collaborator on the ``Auger effect'' in helium. \ As Edoardo
Amaldi has written, an in-depth examination of these works leaves
one struck by their quality: They reveal both a deep knowledge of
the experimental data, even in the minutest detail, and an
uncommon ease, without equal at that time, in the use of the
symmetry properties of the quantum states in order to
qualitatively simplify problems and choose the most suitable
method for their quantitative resolution.

\h In 1932, Majorana published an important paper, (6), on the
non-adiabatic spin-flip of atoms in a magnetic field, which was
later extended by Nobel laureate Rabi in 1937, and by Bloch and
Rabi in 1945. It established the theoretical basis for the
experimental method used to reverse the spin also of neutrons by a
radio-frequency field, a method that is still practiced today, for
example, in all polarized-neutron spectrometers. That paper
contained an independent derivation of the well-known Landau-Zener
formula (1932) for non-adiabatic transition probability. It also
introduced a novel mathematical tool for representing spherical
functions or, rather, for representing spinors by a set of points
on the surface of a sphere (Majorana sphere), attention to which
has been called not long ago by R. Penrose and collaborators [27]
(and by C.Leonardi and coworkers [28]). \ In the {\em present}
Volume the reader will find some additions (or modifications) to
the above-mentioned published articles.

\h However, the most important 1932 paper is that concerning a
relativistic field theory of particles with arbitrary spin, (7).
Around 1932 it was commonly believed that one could write
relativistic quantum equations only in the case of particles with
spin zero or one half. Convinced of the contrary, Majorana
---as we knew since long time from his manuscripts, constituting a part of
the {\em Quaderni} finally published here--- began constructing
suitable quantum-relativistic equations for higher spin values
(one, three-halves, etc.); and he even devised a method for
writing the equation for a generic spin-value. But still he
published nothing\footnote{Starting with 1974, some of us [21]
published and revaluated only few of the pages devoted in its
manuscripts by Majorana to the case of a Dirac-like equation for
the photon (spin-1 case).}, until he discovered that one could
write a single equation to cover an infinite family of particles
of arbitrary spin (even if at that time the known particles could
be counted on one hand). In order to implement his programme with
these ``infinite components'' equations, Majorana invented a
technique for the representation of a group several years before
Eugene Wigner did. And, what is more, Majorana obtained the
infinite-dimensional unitary representations of the Lorentz group
that will be re-discovered by Wigner in his 1939 and 1948 works.
The entire theory was re-invented by Soviet series of articles
from 1948 to 1958, and finally applied by physicists years later.
Sadly, Majorana's initial article remained in the shadows for a
good 34 years until D. Fradkin, informed by E. Amaldi, realized
what Majorana many years earlier had accomplished [26]. All the
scientific material contained in (and in preparation of) this
Majorana's publication is illuminated by the manuscripts constituting
Majorana's Quaderni, a selection of which ---as already mentioned---
will appear in a book [35] that we shall call {\em ``Volume II"}.

\h At the beginning of 1932, as soon as the news of the
Joliot-Curie experiments reached Rome, Majorana understood that
they had discovered the ``neutral proton'' without having realized
it.  Thus, even before the official announcement of the discovery
of the neutron, made soon afterwards by Chadwick, Majorana was
able to explain the structure and stability of light atomic nuclei
with the help of protons and neutrons, antedating in this way also
the pioneering work of D. Ivanenko, as both Segr\'e and Amaldi
have recounted.  Majorana's colleagues remember that even before
Easter he had concluded that protons and neutrons
(indistinguishable with respect to the nuclear interaction) were
bound by the ``exchange forces'' originating from the exchange of
their spatial positions alone (and not also of their spins, as
Heisenberg would propose), so as to produce the alpha particle
(and not the deuteron) as saturated with respect to the binding
energy. Only after Heisenberg had published his own article on the
same problem was Fermi able to persuade Majorana to go for a
six-months period, in 1933, to Leipzig, and meet there his famous
colleague (who will be awarded the Nobel price at the end of that
year); and finally Heisenberg was able to convince Majorana to
publish his results in the paper ``\"Uber die Kerntheorie.''
Actually, Heisenberg had interpreted the nuclear forces in terms
of nucleons exchanging spinless electrons, as if the neutron were
formed in practice by a proton and an electron; whilst Majorana
had simply considered the neutron as a ``neutral proton'': and the
theoretical and experimental consequences were quickly recognized
by Heisenberg. \ Majorana's paper on the stability of nuclei
became soon known to the scientific community ---a rare event, as
we know, for his writings--- thanks to that timely ``propaganda''
made by Heisenberg himself, who on several occasions, when
discussing the ``Heisenberg-Majorana'' exchange forces, used,
rather fairly and generously, to point out more Majorana's than
his own contributions [31]. \ The manuscripts published in the
present book refer also to what Majorana wrote down before having
read Heisenberg's paper. \ Let us seize the present opportunity to
quote two brief passages from Majorana's letters from Leipzig. On
February 14, 1933, he writes his mother (the italics are ours):
``The environment of the physics institute is very nice. I have
good relations with Heisenberg, with Hund, and with everyone else.
{\em I am writing some articles in German. The first one is
already ready} [...]'' [4]. The work that was already ready is,
naturally, the cited one on nuclear forces, which, however,
remained {\em the only paper} in German. Again, in a letter dated
February 18, he tells his father (we italicize): ``{\em I will
publish in German, after having extended it, also my latest
article which appeared in Il Nuovo Cimento}'' [4].

\h But Majorana published nothing more, either in Germany ---where
he had become acquainted, besides with Heisenberg, with other
personalities, including Ehrenfest, Bohr, Weisskopf, Bloch--- or
after his return to Italy: Except for the article (in 1937) of
which we are about to speak. It is therefore of importance to know
that Majorana was engaged in writing other papers: in particular,
he was expanding his article about the infinite-components
equations. \ His research activity during the years 1933-37 is
testified by documents presented below, and particularly by a
number of Majorana's unpublished scientific notes, part of which
are reproduced here: As far as we know, it was focused mainly on
field theory and quantum electrodynamics. \ As we already
mentioned, in 1937 Majorana decided to compete, however, for a
full-professorship (probably with the only desire to have
students); and he was urged to demonstrate that he was still
actively working in theoretical physics: Happily enough, then, he
took from a drawer\footnote{As we said, from the existing
manuscripts it appears that Majorana had formulated also the
essential lines of his paper (9) during the years 1932-33.} his
writing on the symmetric theory of electrons and anti-electrons,
publishing it that same year under the title ``Symmetric theory of
electron and positron.'' This paper ---at present, the most famous
of his, probably--- was initially noticed almost exclusively for
having introduced the Majorana representation of the Dirac
matrices in real form. But its main consequence is that a neutral
fermion can be identical with its anti-particle. Let us stress
that such a theory was rather revolutionary, since it was at
variance with what Dirac had successfully assumed in order to
solve the problem of negative energy states in quantum field
theory. With rare daring, Majorana suggested that neutrinos, which
had just been postulated by Pauli and Fermi to explain puzzling
features of radioactive beta decay, could be particles of this
type. This would enable the neutrino, for instance, to have mass;
which may have a bearing on the phenomena of neutrino
oscillations, later postulated by Pontecorvo.

\h It may be stressed that, exactly as in the case of other
writings of his, the ``Majorana neutrino'' too started to gain
prominence only decades later, beginning in the 1950s; and
nowadays expressions like Majorana spinors, Majorana mass, and
even ``majorons'', are fashionable: It is moreover well-known that
many experiments are currently devoted all the world over to check
whether the neutrinos are of the Dirac or the Majorana type. \ We
have already said that the material itself, {\em published} by
Majorana (but still little known, despite it all), constitutes a
potential gold-mine for physics. \ Many years ago, for example,
Bruno Touschek noticed that the present article implicitly
contains also what he called the theory of the ``Majorana
oscillator,'' described by the simple equation \
$q+\omega^2q=\varepsilon \ \delta (t)$, \ where $\varepsilon$ is a
constant and $\delta$ the Dirac function [4]. According to
Touschek, the properties of the Majorana oscillator are very
interesting, especially in connection with its energy spectrum:
But no literature seems to exist yet on it.

\

\section*{4. An account of the unpublished manuscripts}

\noi The largest part of the Majorana's work was left unpublished.
Even if the most important manuscripts have gone probably lost, we
are now in possession of: \ (i) his M.Sc thesis on ``The quantum
theory of radioactive nuclei''; \ (ii) 5 notebooks (the {\em
Volumetti}); \ 18 booklets (the {\em Quaderni}); \ (iii) 12
folders with loose papers, and \ (iv) the set of his lecture notes
for the course on theoretical physics given by him at the
University of Naples. With the collaboration of E. Amaldi, all
these manuscripts were deposited by Luciano Majorana (Ettore's
brother) at the Domus Galilaeana of Pisa, Italy.  An analysis of
those manuscripts allowed us to ascertain that they, except for
the lectures notes, appear to have been written approximately by
1933 (even the essentials of his last article, which Majorana
proceeded to publish, as we already know, in 1937, seem to have
been ready by 1933, the year in which the discovery of the
positron was confirmed).  Besides the material deposited at the
Domus Galilaeana, we have been in possession of: \ (v) a series of
34 letters written by Majorana between March 17, 1931, and
November 16, 1937, in reply to his uncle Quirino ---a renowned
experimental physicist and a former president of the Italian
Physical Society--- who had been pressing Majorana for help in the
theoretical explanation of his experiments:\footnote{In the past,
one of us (E.R.) have been able to publish only short passages of
them, since they are rather technical: cf. Ref.[4].} such letters
have been recently deposited at the Bologna University, and have
been published in their entirety by G. Dragoni [8]. They confirm
that Majorana was deeply knowledgeable even about experimental
details. Moreover, Ettore's sister, Maria, recalled that, even in
those years, Majorana ---who had reduced his visits to Fermi's
Institute, starting from the beginning of 1934 (that is, just
after his return from Leipzig)--- continued to study and work at
home many hours during the day and at night. Did he continue to
dedicate himself to physics? From one of those letters of his to
Quirino, dated January 16, 1936, we find a first answer, because
we get to learn that Majorana had been occupied ``since some time,
with quantum electrodynamics;'' knowing Majorana's love for
understatements, this no doubt means that during 1935 Majorana had
performed profound research at least in the field of quantum
electrodynamics.

\h This seems to be confirmed by a recently retrieved text,
written by Majorana in French [25], where the author dealt with a
peculiar topic in quantum electrodynamics. It is instructive, as
to that topic, just to quote directly from the Majorana's paper:

\

\h $\ll$Let us consider a system of $p$ electrons and set the
following assumptions: 1) the interaction between the particles be
sufficiently small, allowing to speak about individual quantum
states, so that one may regard the quantum numbers defining the
configuration of the system as good quantum numbers; 2) any
electron has a number $n > p$ of inner energetic levels, while any
other level has a much greater energy. One deduces that the states
of the system as a whole may be divided into two classes. The
first one is composed of those configurations for which all the
electrons belong to one of the inner states. Instead, the second
one is formed by those configurations in which at least one
electron belongs to a higher level not included in the
above-mentioned $n$ levels. We shall also assume that it is
possible, with a sufficient degree of approximation, to neglect
the interaction between the states of the two classes. In other
words, we will neglect the matrix elements of the energy
corresponding to the coupling of different classes, so that we may
consider the motion of the $p$ particles, in the $n$ inner states,
as if only these states existed. Our aim becomes, then,
translating this problem into that of the motion of $n-p$
particles in the same states, such new particles representing the
holes, according to the Pauli principle.$\gg$

\

\noi Majorana, thus, applied the formalism of field quantization
to Dirac's hole theory, obtaining a general expression for the QED
Hamiltonian in terms of anticommuting ``hole quantities''. Let us
point out that, in justifying the use of anticommutators for
fermionic variables, Majorana commented that such a use ``cannot
be justified on general grounds, but only by the particular form
of the Hamiltonian. In fact, we may verify that the equations of
motion are better satisfied by these relations than by the
Heisenberg ones.'' \  In the second (and third) part of the same
manuscript, Majorana took into consideration also a reformulation
of QED in terms of a photon wavefunction: a topic that was
particularly studied in his {\em Quaderni} (and is reproduced
here). Majorana, indeed, reformulated quantum electrodynamics by
introducing a real-valued wave function for the photon,
corresponding only to directly observable degrees of freedom.

\h In some other manuscripts, probably prepared for a seminar at
the Naples University in 1938 [24], Majorana set forth a physical
interpretation of quantum mechanics, that anticipated of several
years the Feynman approach in terms of path integrals. The
starting point in Majorana's notes was to search for a meaningful
and clear formulation of the concept of quantum state. Afterwards,
the crucial point in the Feynman formulation of quantum mechanics
(namely, that of considering not only the paths corresponding to
classical trajectories, but all the possible paths joining an
initial point with the final point) was really introduced by
Majorana, after a discussion about an interesting example of
harmonic oscillator. Let us also emphasize the key role played by
the symmetry properties of the physical system in the Majorana
analysis; a feature quite common in this author's papers.

\h Do any other unpublished scientific manuscripts of Majorana
exist?  The question, raised by his answer to Quirino and by his
letters from Leipzig to his family, becomes of greater importance
when one reads also his letters addressed to the National Research
Council of Italy (CNR) during that period. In the first one (dated
January 21, 1933), Majorana asserts: ``At the moment, I am
occupied with the elaboration of a theory for the description of
arbitrary-spin particles that I began in Italy and of which I gave
a summary notice in Il Nuovo Cimento [...]'' [4].  In the second
one (dated March 3, 1933) he even declares, referring to the same
work: ``I have sent an article on nuclear theory to Zeitschrift
f\"ur Physik. I have the manuscript of a new theory on elementary
particles ready, and will send it to the same journal in a few
days'' [4]. Considering that the article described above as a
``summary notice'' of a new theory was already of a very high
level, one can imagine how interesting it would be to discover a
copy of its final version, which went unpublished. [Is it still,
perhaps, in the Zeitschrift f\"ur Physik archives? Our search
ended till now in failure].

\h A few of Majorana's other ideas, which did not remain concealed
in his own mind, have survived in the memories of his colleagues.
One such reminiscence we owe to Gian Carlo Wick.  Writing from
Pisa on October 16, 1978, he recalls: ``The scientific contact
[between Ettore and me], mentioned by Segr\'e, happened in Rome on
the occasion of the `A. Volta Congress' (long before Majorana's
sojourn in Leipzig).  The conversation took place in Heitler's
company at a restaurant, and therefore without a blackboard [...];
but even in the absence of details, what Majorana described in
words was a `relativistic theory of charged particles of zero spin
based on the idea of field quantization' (second quantization).
When much later I saw Pauli and Weisskopf's article [{\it Helv.
Phys. Acta} {\bf 7} (1934) 709], I remained absolutely convinced
that what Majorana had discussed was the same thing [...]'' [4].

\

\section*{5. Teaching theoretical physics}

\noi As we have seen, Majorana did significantly contribute to
theoretical research which was among the frontier topics in the
1930s, and, indeed, in the following decades. However, he deeply
thought also on the basics, and applications, of quantum
mechanics: and Majorana's lectures on theoretical physics forward
an evidence of this work of his.

\

\h As realized only recently [32], Majorana revealed a genuine interest in advanced
physics teaching, starting from 1933, just after he obtained, at the end of 1932, the
degree of ``libero docente'' (analogous to the German ``privatdozent'' title). As allowed
by that degree, he requested to be allowed to give three subsequent annual free courses
at the University of Rome, between 1933 and 1937, as testified by the lecture programs
proposed by him and still present in the Rome University's archives. Such documents too
refer to a period of time that had been regarded by his colleagues as Majorana's ``gloomy
years''. Although it seems that Majorana never delivered such three courses, probably due
to lack of appropriate students, the chosen topics to be lectured appear very interesting
and informative.

\h The first course (academic year 1933-34) proposed by Majorana
was that of Mathematical Methods of Quantum Mechanics.\footnote{The program
for it contained the following topics: \ (i) Unitary geometry. Linear
transformations. Hermitian operators. Unitary transformations. Eigenvalues
and eigenvectors; \ (ii) Phase space and the quantum of action. Modifications
of classical kinematics. General framework of quantum mechanics; \  (iii)
Hamiltonians which are invariant under a transformation group.
Transformations as complex quantities. Non compatible systems.
Representations of finite or continuous groups; \ (iv) General
elements on abstract groups. Representation theorems. The group of
spatial rotations. Symmetric groups of permutations and other
finite groups; \ (v) Properties of the systems endowed with spherical
symmetry. Orbital and intrinsic momenta. Theory of the rigid
rotator; \ (vi) Systems with identical particles. Fermi and
Bose-Einstein statistics. Symmetries of the eigenfunctions in the
center-of-mass frames; \ (vii) Lorentz group and spinor
calculus. Applications to the relativistic theory of the
elementary particles.}

\h The second proposed course (academic year 1935-36) was instead
that of Mathematical Methods of Atomic Physics.\footnote{The
corresponding subjects were: \ Matrix calculus. Phase space and
the correspondence principle. Minimal statistical sets or
elementary cells. Elements of quantum dynamics. Statistical
theories. General definition of symmetry problems. Representations
of groups. Complex atomic spectra. Kinematics of the rigid body.
Diatomic and polyatomic molecules. Relativistic theory of the
electron and the foundations of electrodynamics. Hyperfine
structures and alternating bands. Elements of nuclear physics.}

\h Finally, the requested third course (academic year 1936-37) was that of
Quantum Electrodynamics.\footnote{The main topics were: \ Relativistic
theory of the electron. Quantization procedures.
Field quantities defined by commutability and anticommutability
laws. Their kinematical equivalence with sets with an undetermined
number of objects obeying the Bose-Einstein or Fermi statistics,
respectively. Dynamical equivalence. Quantization of the
Maxwell-Dirac equations. Study of the relativistic invariance. The
positive electron and the symmetry of charges. Several
applications of the theory. Radiation and scattering processes.
Creation and annihilation of opposite charges. Collisions of fast
electrons.}

\h Majorana could actually lecture on theoretical physics only in 1938 when, as recalled
above, he obtained his position as a full-professor in Naples. He gave his lectures
starting on January 13 and ending with his disappearance (March 26), but his activity was
intense, and his interest for teaching very high. For the benefit of his students, and
perhaps also for writing down a book, he prepared careful lecture notes [17,18]. A recent
analysis [34] showed that Majorana's 1938 course was very innovative for that time, and
this has been confirmed by the retrieval (on September 2004) of a faithful transcription
of the whole set of Majorana's lecture notes (the so-called ``Moreno document'')
comprising the 6 lectures not included in the original collection [19].

\h The first part of his course on theoretical physics dealt with
the phenomenology of the atomic physics and its interpretation in
the framework of the old Bohr-Sommerfeld quantum theory. This
part presents a strict analogy with the course given by Fermi in
Rome (1927-28), attended by Majorana when a student. \ The second
part started, instead, with the classical radiation theory,
reporting explicit solutions to the Maxwell equations, scattering
of the solar light and some other applications. It then continued
with the theory of Relativity: after the presentation of the
corresponding phenomenology, a complete discussion of the
mathematical formalism required by that theory was given, ending
with some applications as the relativistic dynamics of the
electron. Then, it followed a discussion of important effects for
the interpretation of quantum mechanics, such as photoelectric
effect, Thomson scattering, Compton effects and the Franck-Hertz
experiment. \ The last part of the course, more mathematical in
nature, treated explicitly quantum mechanics, both in the
Schr\"odinger and in the Heisenberg formulations. This part did
not follow the Fermi approach, but rather referred to personal
previous studies, getting also inspiration from Weyl's book on
group theory and quantum mechanics.

\

\section*{6. A brief sketch of ``Volume I" (which includes\\
Majorana's ``study" notes)}

\noi In Volume I we have reproduced, and translated, Majorana's
{\em Volumetti}: that is, his {\em study} notes, written in Rome
between 1927 and 1932. Each one of those neatly organized
booklets, prefaced by a table of contents, consisted in about
100$-$150 sequentially numbered pages, while a date, penned on its
first blank, recorded the approximate time during which it was
completed. Each {\em Volumetto} was written during a period of
about one year. The contents of those notebooks range from typical
topics covered in academic courses to topics at the frontiers of
research: despite this unevenness in the level of sophistication,
the style is never obvious. As an example, we can recall
Majorana's study of the shift in the melting point of a substance
when it is placed in a magnetic field, or his examination of heat
propagation using the ``cricket simile.'' As to frontier research
arguments, we can recall two examples: the study of
quasi-stationary states, anticipating Fano's theory, and the
already mentioned Fermi's theory of atoms, reporting analytic
solutions of the Thomas-Fermi equation with appropriate boundary
conditions in terms of simple quadratures. He also treated
therein, in a lucid and original manner, contemporary physics
topics such as Fermi's explanation of the electromagnetic mass of
the electron, the Dirac equation with its applications, and the
Lorentz group.

\h Just to give a very short account of the interesting material
present in the {\em Volumetti}, let us point out the following.

\h First of all, we already mentioned that in 1928, when Majorana
was starting to collaborate (still as a University student) with
the Fermi group in Rome, he already revealed his outstanding
ability in solving involved mathematical problems in original and
clear ways, by obtaining an analytical series-solution of the
Thomas-Fermi equation. Let us recall once more that his whole work
on this topic was written in some loose sheets, and then
diligently transcribed by the author himself in his {\em
Volumetti}: so that it is contained in Volume I. From those pages,
it appears evident also the contribution given by Majorana to the
relevant statistical model, anticipating some important results
found later on by leading specialists. As to the major Majorana's
finding (namely, his methods of solutions of that equation), let
us stress that it remained completely unknown until very recent
times, at the extent that the physicist community ignored that the
non-linear differential equations, relevant for atoms and for
other systems too, can be solved semi-analytically (see Sect. 7 of
{\em Volumetto II}). Indeed, a noticeable property of the method
invented by Majorana for solving the Thomas-Fermi equation, is
that it may be easily generalized, and may then be applied to a
large class of particular differential equations. Several
generalizations of his method for atoms were proposed by Majorana
himself: they have been rediscovered only many years later.  For
example, in Sect. 16 of {\em Volumetto II}, Majorana studied the
problem of an atom in a weak external electric field, that is, the
problem of atomic polarizability, and obtained an expression for
the electric dipole moment for a (neutral or arbitrarily ionized)
atom. Furthermore, he also started applying the statistical method
to molecules, rather than single atoms, by studying the case of a
diatomic molecule with identical nuclei (see Sect. 12 of {\em
Volumetto II}). Finally, our author considered the second
approximation for the potential inside the atom, beyond the
Thomas-Fermi approximation, by generalizing the statistical model
of neutral atoms to those ionized $n$ times, the case $n = 0$
included (see Sect. 15 of {\em Volumetto II}). As recently pointed
out by one of us (S.E.) [23], the approach used by Majorana to
this end is rather similar to the one now adopted in the
renormalization of physical quantities, in modern gauge theories.

\h As well documented, Majorana was among the first ones to study
nuclear physics in Rome (we already know that in 1929 he defended
a M.Sc. thesis on such a subject). But he continued to do research
on similar topics for several years, till his famous 1933 theory
of nuclear exchange-forces. \ For $(\alpha,p)$ reactions on light
nuclei, whose experimental results had been interpreted by
Chadwick and Gamov, in 1930 Majorana elaborated a dynamical theory
(in Sect. 28 of {\em Volumetto IV}) by describing the energy
states associated with the superposition of a continuous spectrum
and one discrete level [33]. \ Actually, Majorana provided a
complete theory for the artificial disintegration of nuclei
bombarded by $\alpha$ particles (with and without $\alpha$
absorption). He approached this question by considering the
simplest case, with a single unstable state of a nucleus and an
$\alpha$ particle, which spontaneously decays by emitting an
$\alpha$ particle or a proton. The explicit expression for the
total cross-section was also given, rendering his approach
accessible to experimental checks. Let us emphasize that the
peculiarity of Majorana's theory was the introduction of
quasi-stationary states, which were considered by U. Fano in 1935
(in a quite different context), and widely used in condensed
matter physics about 20 years later.

\h In Sect. 30 of {\em Volumetto II}, Majorana made an attempt to
find a relation between the fundamental constants $e, \, h, \, c$.
The interest of this work resides less in the particular
mechanical model adopted by Majorana (which led, indeed, to the
result $e^2 \simeq h c$ far from the true value, as noticed by the
author himself), than in the adopted interpretation of the
electromagnetic interaction, in terms of particle exchange.
Namely, the space around charged particles was regarded as
quantized, and electrons interacted by exchanging particles;
Majorana's interpretation substantially coincides with that
introduced by Feynman in quantum electrodynamics after more than a
decade, when the space surrounding charged particles will be
identified with the QED vacuum, while the exchanged particles will
be assumed to be photons.

\h Finally, one cannot forget the pages contained in {\em
Volumetti III} and {\em V} on group theory, where Majorana showed
in detail the relationship existing between the representations of
the Lorentz group and the matrices of the (special) unitary group
in two dimensions. In those pages, aimed also at extending Dirac's
approach, Majorana deduced the {\it explicit} form of the
transformations of every bilinear quantities in the spinor fields.
Certainly, the most important result achieved by Majorana on this
subject is his discovery of the {\it infinite-dimensional} unitary
representations of the Lorentz group: He set forth the {\it
explicit} form of them too (see Sect. 8 of {\em Volumetto V},
besides his published article (7).) \ We already recalled that
such representations were rediscovered by Wigner only in 1939 and
1948, and later on, in the years 1948-1958, eventually studied by
many authors. People like van der Waerden recognized the
importance, also mathematical, of such a Majorana's result, but,
as we know, it remained unnoticed till the above-quoted 1966
Fradkin's article.

\

\section*{7. About Majorana's research notes (``Volume II")}

\noi The material reproduced in Volume I was a paragon of order,
conciseness, essentiality and originality. So much so that those
notebooks can be partially regarded as an innovative text of
theoretical physics, even after about eighty years; besides being
another gold-mine of theoretical, physical, and mathematical ideas
and hints, stimulating and useful for modern research too.

\h  But Majorana's most remarkable scientific manuscripts
---namely, his {\em research} notes--- are represented by a
host of loose papers and by the {\em Quaderni}: and what we called
``Volume II" will reproduce[35] a selection of the latter. \ But
the manuscripts with Majorana's research notes, at variance with
the {\em Volumetti}, do rarely contain any introductions or verbal
explanations.

\h The topics faced in the {\em Quaderni} range
from classical physics to quantum field theory, and comprise the study of a
number of applications for atomic, molecular and nuclear physics. \
Particular attention was reserved to the Dirac
theory and its generalizations, and to quantum electrodynamics.

\h  The Dirac equation describing spin-1/2 particles was mostly
considered by Majorana in a {\it lagrangian framework} (in
general, the canonical formalism was adopted), obtained from a
least action principle: as shown in a Chapter of the coming
Volume [35]. After an interesting preliminary study of the problem of
the vibrating string, where Majorana obtained a (classical)
Dirac-like equation for a two-component field, he then went on to
consider a semiclassical relativistic theory for the electron,
within which the Klein-Gordon and the Dirac equations were deduced
starting from a semiclassical Hamilton-Jacobi equation.
Subsequently, the field equations and their properties were
considered in detail, and the quantization of the (free) Dirac
field discussed by means of the standard formalism, with the use
of annihilation and creation operators. Then, the electromagnetic
interaction was introduced into the Dirac equation, and the
superposition of the Dirac and Maxwell fields were studied in a
very personal and original way, obtaining the expression for the
quantized Hamiltonian of the interacting system after a
normal-mode decomposition.

\h Real (rather than complex) Dirac fields, published by Majorana
in his famous paper, (9), on the symmetrical theory of electrons
and positrons, were considered in the Quaderni in various
places (see Quaderno 3, page 3, in ref.[35]), by two slightly different
formalisms, namely, by different decompositions of the field. The
introduction of the electromagnetic interaction was performed in a
quite characteristic manner, and he then obtained an {\it
explicit} expression for the total angular momentum, carried by
the real Dirac field, starting from the Hamiltonian.

\h  Some work, as well, at the basis of Majorana's important paper
(7), can be found in the Quaderni under consideration (as it will be
seen, e.g., in a Section of ref.[35]).  We have already seen above,
when analyzing the works published by Majorana, that Majorana in
1932 constructed Dirac-like equations for spin one, three-halves,
two, etc. (discovering also the method, later published by Pauli
\& Fiertz, for writing down a quantum-relativistic equation for a
generic spin-value). Indeed, in the Quaderni reproduced
here, our author, starting from the usual Dirac equation for a
4-component spinor, obtains {\it explicit} expressions for the
Dirac matrices in the cases, for instance, of 6-component and
16-component spinors. Interestingly enough, at the end of his
discussion, Majorana also treats the case of spinors with an {\it
odd} number of components, namely, of a 5-component field.

\h  With regard to quantum electrodynamics too, Majorana dealt
with it in a lagrangian and hamiltonian framework, by use of a
least action principle. As it is {\it now} currently done, the
electromagnetic field was decomposed in plane-wave operators, and
its properties were studied within a {\it full Lorentz-invariant
formalism} by employing group-theoretical arguments. Explicit
expressions for the quantized Hamiltonian, the creation and
annihilation operators for the photons, as well as the angular
momentum operator, were deduced in several different bases, along
with the appropriate commutation relations. Even leaving aside,
for a moment, the scientific value those results had especially at
the time when Majorana got them, such manuscripts bear a certain
importance from the historical point of view too: They indicate
Majorana's tendency to tackle with topics of that kind, nearer to
Heisenberg, Born, Jordan and Klein's, than to Fermi's.

\h  As we were saying, and as already pointed out in previous
literature [21], in the Quaderni one can find also various
studies, inspired to an Oppenheimer's idea, aimed at describing
the electromagnetic field within a Dirac-like formalism. Actually,
Majorana was interested in describing the properties of the
electromagnetic field in terms of a real wavefunction for the
photon (see Quaderno 2, page 101a, and Quaderno 17, page 142), an approach
that went well beyond the work of contemporary authors. \ Other
noticeable Majorana's investigations concerned the introduction of
an {\it intrinsic} time delay, regarded as a universal constant,
into the expressions for electromagnetic retarded fields (see Quaderno 5, page 65);
 or studies on the modification of Maxwell's
equations in the presence of magnetic monopoles (see Quaderno 3, page 163).

\h  Besides purely theoretical work in quantum electrodynamics, some
applications as well were carefully investigated by Majorana. This is the
case of the free electron scattering (investigated in Quaderno 17, page 133, as
reported in ref.[35]), where Majorana gave an {\it explicit} expression
for the transition probability, and the coherent scattering, of
bound electrons (see Quaderno 17, page 142b). Several other
scattering processes were also analyzed (e.g., in a Chapter of ref.{[35])
within the framework of perturbation theory, by the adoption of
Dirac's or of Born's method.

\h  As remarked above, the contribution by Majorana in nuclear
physics, which became most known to the scientific community of
his times, is his theory in which nuclei are formed by protons and
neutrons, bound by an exchange-force of a particular kind (which
corrected Heisenberg's model). In the Quaderni (cf., e.g., a Chapter
of ref.[35]), several pages were devoted to analyze
possible forms of the nucleon potential inside a given nucleus,
determining the interaction between neutrons and protons. Although
general nuclei were often taken into consideration, particular
care was given by the author to light nuclei (deuteron,
$\alpha$-particle, etc.). As it will be clearer from what will appear
published in ref.[35], the studies performed by Majorana were,
at the same time, preliminary studies, and generalizations of what
published by him in his well-known publication (8), thus revealing
a {\em very rich} and personal way of thinking. Notice also that,
before having understood and thought all what led him to the
mentioned paper (8), Majorana had seriously attempted at
constructing a {\em relativistic field theory for nuclei} as
composed of scalar particles (see Quaderno 2, page 86), arriving at
a characteristic description of the transitions between different
nuclei.

\h Other topics in nuclear physics were broached by our author
(and were present in the {\em Volumetti} too): We shall only
mention, here, the study of the energy loss of $\beta$ particles
when passing through a medium, when he deduced the Thomson formula
by classical arguments. Such a work too might a priori be of
interest for a correct historical reconstruction, when confronted
with the very important theory on nuclear $\beta$ decay elaborated
by Fermi in 1934.

\h The largest part of the Quaderni is devoted, however, to
atomic physics (see, e.g., ref.[35]), in agreement with the
circumstance that it was the main research topic tackled by the
Fermi group in Rome in the years 1928-1933. Indeed, also the
articles published by Majorana in those years deal with such a
subject; and echoes of those publications can be found, of course,
in the Quaderni: Showing that, especially in the
case of the article (5) on the incomplete $P^\prime$ triplets,
some {\em interesting} material did not appear in the published
papers.

\h Several expressions for the wavefunctions and the different
energy levels of two-electron atoms (and, in particular, of
helium) were discovered by Majorana, mainly in the framework of a
variational method aimed at solving the relevant Schr\"odinger
equation. Numerical values for the corresponding energy terms were
normally summarized by Majorana in large tables. Some approximate
expressions were also obtained by him
for three-electron atoms (and, in particular, for lithium), and
for alkali; including the effect of polarization forces in
hydrogen-like atoms.

\h In the Quaderni, the problem of the hyperfine
structure of the energy spectra of complex atoms was moreover
investigated in some detail, revealing[35] the careful attention paid
by Majorana to the existing literature.  The generalization, for a
{\em non-coulombian} atomic field, of the Land\`e formula for the
hyperfine splitting was also performed by Majorana, together with
a {\em relativistic} formula for the Rydberg corrections of the
hyperfine structures. Such a detailed study developed by Majorana
constituted the basis of what discussed by Fermi and Segr\`e in a
well-known 1933 paper of theirs on this topic, as acknowledged by
those authors themselves.

\h A small part of the Quaderni was devoted to various
problems of molecular physics (see ref.[35]). Majorana
studied in some detail, for example, the helium molecule, and,
then, considered the general theory of the vibrational modes in
molecules, with particular reference to the molecule of acetylene,
$C_2H_2$ (which possesses peculiar geometric properties).

\h  Rather important are some other pages (see Quaderno 8, page 14
and page 46; and Quaderno 6, page 8), where the author considered the
problem of ferromagnetism in the framework of Heisenberg's model
with exchange interactions. However, Majorana's approach in this
study was, as always, {\it original}, since it did follow neither
Heisenberg's, nor the subsequent van Vleck's formulation in terms
of a spin Hamiltonian. By using statistical arguments, instead,
Majorana evaluated the magnetization (with respect to the
saturation value) of the ferromagnetic system when an external
magnetic field acts on it, and the phenomenon of spontaneous
magnetization. Several examples of ferromagnetic materials, with
different geometries, were by him analyzed as well.

\h  A number of other interesting questions, even dealing with
topics that Majorana had encountered during his academic studies
at Rome University (as witnessed by a couple of Chapters of
ref.[35]),
can be found in the Quaderni. This is the case, for
example, of the electromagnetic and {\it electrostatic} mass of
the electron (a problem that was considered by Fermi in one of his
1924 known papers); or of his studies on tensor calculus,
following his teacher Levi-Civita. Here, we cannot go on in their
discussion, however, our aim being that of drawing the attention
of the reader to a few specific points only. It is left to the
reader's patience the discovery of the large number of exceedingly
interesting and important studies that were faced by Majorana, and
written by him in the Quaderni [35].

\

\cent{\Large{\bf APPENDIX}}

\section*{8. Appendix: About the format of Majorana's scientific manuscripts}

\noi As it is clear from what discussed above, Majorana used to
put on paper the results of his studies following different
behaviors, depending on his opinion about the value of the results
themselves. The method used by Majorana for composing his written
notes was sometimes the following. When he was investigating a
certain subject, he reported his results only in a {\em Quaderno}.
Subsequently, if, after further research on that topic,
Majorana reached a simpler and conclusive (in his opinion) result,
he then reported the final details also in a {\em Volumetto}.
Therefore, in his preliminary notes we find basically mere
calculations, and only in some rare cases an elaborate text,
explaining his calculations, can be found in the {\em
Quaderni}. In other words, a clear verbal exposition of the
topic can sometimes be found only in the {\em Volumetti}.

\h The eighteen Quaderni deposited at the Domus Galilaeana
in Pisa
are booklets approximately of 15cm$\times$21cm, endowed with a
black cover and a red external boundary, as it was common in Italy
before the Second World War. Each booklet is composed of about 200
pages, for a total of about 2800 pages. Rarely, some pages were
teared off (by Majorana himself), while blank pages in each
Quaderno are often present. In few booklets, extra pages written
by the author were put in.

\h An original numbering of the pages is present only in {\em
Quaderno} 1 (on the central top of each page). However, all the
Quaderni present a non original numbering (written in red
ink) at the left top corner of their odd pages. Blank pages too
have been always numbered. Interestingly enough, even if an
original numbering by Majorana in general is not present,
nevertheless sometimes in a Quaderno it appears an original
reference to some pages of that same booklet. Some other strange
cross-references, not easily understandable for us, do appear
in various booklets.

\h As it was evident also from a previous catalog of the
unpublished manuscripts, prepared long ago by Baldo, Mignani and
Recami [14], often in the Quaderni the material regarding the same
subject was not written in a sequential, logical order: In
some cases, it even appears in the reverse order.

\h The major part of the Quaderni contains calculations
without any explicit explanations; even if, in few cases, an elaborate
text is fortunately present.

\h At variance with what happens for the {\em Volumetti}, in the
Quaderni no dates appear, except for the {\em Quaderni} 16
(of 1929-30), 17 (star\-ted on 20 June 1932) and, probably, 7
(approximately of 1928). Therefore, the actual dates of composition
of the manuscripts may be inferred only from a detailed comparison of
the topics, studied therein, with what is present in the {\em
Volumetti} and in the published literature, including Majorana's
published papers. Some additional information comes from some
cross-references explicitly penned by the author himself,
referring either to his {\em Quaderni} or to his {\em Volumetti}.
In few cases, references to the some of the existing literature
are explicitly introduced by Majorana.

\h No consequential or time order is present in the
Quaderni, but nevertheless we have been able in ref.[35]
to group the topics into four (large) Parts. In any case,
in Volume II it will be reported, in a second Index, the
complete list of the subjects present in the eighteen
Quaderni.

\h We have done a major effort in carefully checking and typing
all equations and tables; and, even more, in writing down a brief
presentation of the argument exploited in each subsection. In
addition, we have inserted among Majorana's calculations a minimum
number of words, when he had left his formalism without any text,
trying to facilitate the reading of Majorana's research notebooks,
but limiting as much as possible the insertion of any personal
comments of ours. Our hope is, by ref.[35], to render the
intellectual treasure, contained in the {\em Quaderni}, accessible
for the first time to the widest audience.

\h At the end of this paper, we attach a short Bibliography. Far
from being exhaustive, it just provides references
about the topics touched upon in the present article.


{}\vfill

\section*{9. Acknowledgements}

\

\noi This work has been done in collaboration with A. van der Merwe
and R. Battiston.  \  We are moreover indebted, for their
kind help, to C. Segnini, the former
curator of the Domus Galileana in Pisa, and to the previous
curators and directors too. \ Thanks are moreover due to A. Drago, A.
De Gregorio, E. Giannetto, E. Majorana jr., and F. Majorana for
valuable cooperation along the years.

\

\

\section*{10. \Large \bf BIBLIOGRAPHY} %
\addcontentsline{toc}{chapter}{{${}$ \newline \noi ${}$
\hspace{-0.7truecm} {\bf Bibliography}}} %

\


\noi Biographical papers, written by witnesses who knew Ettore
Majorana personally, are the following:

\begin{enumerate}
\item[{[1]}] E. Amaldi, {\em La Vita e l'Opera di Ettore Majorana}
(Accademia dei Lincei; Rome, 1966). \ Cf. also E. Amaldi, ``Ettore
Majorana: Man and Scientist'', in {\em Strong and Weak
Interactions}, edited by A. Zichichi (Academic Press; New York,
1966); \ ``Ettore Majorana, a cinquant'anni dalla sua scomparsa,''
in {\em Nuovo Saggiatore} {\bf 4} (1988) 13; \ ``From the
discovery of the neutron to the discovery of nuclear fission'', in
{\em Physics Reports} {\bf 111} (1984) pp.1--322.

\item[{[2]}] B. Pontecorvo, {\em Fermi e la fisica moderna}
(Editori Riuniti; Rome, 1972); \ and \ in {\em Proceedings of the
International Conference on the History of Particle Physics,
Paris, July 1982} in {\em Physique} {\bf 43} (1982);

\item[{[3]}] E. Segr\'e, {\em Enrico Fermi, Physicist} (University
of Chicago Press; Chicago, 1970); \ {\em A Mind Always in Motion}
(University of California Press; Berkeley, 1993).
\end{enumerate}

\noi Accurate biographical information, completed by the
reproduction of very many documents, has to be found in the
following book (where almost all the relevant documents existing
by 2002 ---discovered or collected by that author--- appeared for
the first time):

\begin{enumerate}
\item[{[4]}] E. Recami, {\em Il caso Majorana: Epistolario,
Documenti, Testimonianze}, 2nd ed. (Mondadori; Milan, 1991); 4th
ed. (Di Renzo; Rome, 2002), pp.1-273.
\end{enumerate}

\noi See also:

\begin{enumerate}
\item[{[5]}] E. Recami, ``Ricordo di Ettore Majorana a
sessant'anni dalla sua scomparsa: l'opera scientifica edita e
inedita,'' in {\em Quaderni di Storia della Fisica (della
Societ\`a Italiana di Fisica)} {\bf 5} (1999) 19-68, arXive:
physics/9810023, and refs. therein. \ Cf. also: E.Recami, ``New
Evidences about the Disappearance of the Physicist Ettore
Majorana", in {\em Scientia} {\bf 110} (1975) 577-598; and in {\em
Tachyons, Monopoles, and Related Topics}, ed. by E.Recami
(North-Holland; Amsterdam, 1978), the Appendix, pp.267-277; as
well as the recent bilingual e-print
arXiv:0708.2855[physics.hist-ph].

\item[{[6]}] F. Cordella, A. De Gregorio and F. Sebastiani, {\em
Enrico Fermi. Gli anni italiani} (Editori Riuniti; Rome, 2001);

\item[{[7]}] S. Esposito, ``Fleeting genius,'' in {\em Physics
World} {\bf 19} (2006) 34; E. Recami, ``Majorana: His scientific
and human personality,'' in {\em Proceedings of the International
Conference on Ettore Majorana's legacy and the Physics of the XXI
Century}, PoS(EMC2006)016 (SISSA; Trieste, 2006);

\item[{[8]}] G. Dragoni (ed.), {\em Ettore e Quirino Majorana tra
Fisica Teorica e Sperimentale},
(Pubblicazioni Scientifiche CNR; Rome, in press).
\end{enumerate}

\noi Scientific published articles by Majorana have been discussed
and/or translated into English in the following papers:

\begin{enumerate}
\item[{[9]}] ``Ettore Majorana -- On nuclear theory [Zeits. f.
Physik {\bf 82} (1933) 137]'': English translation appeared in
Part 2 of the book by D.M. Brink, {\em Nuclear Forces} (Pergamon
Press; Oxford, 1965);

\item[{[10]}] ``Ettore Majorana -- Relativistic theory of
particles with arbitrary intrinsic angular momentum [Nuovo Cimento
{\bf 9} (1932) 335]'': English translation by C.A. Orzalesi, {\em
Tecnical Report no.792, Feb. 1968 (Dept. of Phys. and Astr.; Univ.
of Maryland; College Park)}.

\item[{[11]}] ``Ettore Majorana -- Symmetrical theory of the
electron and the positron [Nuovo Cimento {\bf 14} (1937) 171]'':
English translation by D.A. Sinclair, {\em Technical Translation
no.TT-542 (National Research Council of Canada; 1975)}.

\item[{[12]}] ``Ettore Majorana -- A symmetric theory of electrons
and positrons [Nuovo Cimento {\bf 14} (1937) 171]'': English
translation by L. Maiani, appeared in {\em Soryushiron Kenkyu}
{\bf 63} (1981) 149.

\item[{[13]}] {\em Ettore Majorana -- Scientific Papers}, edited
by G.F. Bassani (S.I.F., Bologna, and Springer, Berlin; 2006).
\end{enumerate}

\noi A preliminary catalog of the unpublished papers by Majorana
did first appear in:

\begin{enumerate}
\item[{[14]}] M. Baldo, R. Mignani, and E. Recami, ``Catalogo dei
manoscritti scientifici inediti di E. Majorana,'' in Ref.{[17]};
as well as in Ref.{[5]} [arXive: physics/9810023].
\end{enumerate}

\noi The English translation of the {\em Volumetti} appeared as

\begin{enumerate}
\item[{[15]}] S. Esposito, E. Majorana Jr., A. van der Merwe and
E. Recami (editors), {\em Ettore Majorana -- Notes on Theoretical
Physics} (Kluwer Acad. Pub.; Dordrecht and Boston, 2003).
\end{enumerate}

\noi The original Italian version, instead, has been successively
published in:

\begin{enumerate}
\item[{[16]}] S. Esposito and E. Recami (editors), {\em Ettore
Majorana -- Appunti Inediti di Fisica Teorica} (Zanichelli;
Bologna, 2006).
\end{enumerate}

\noi The anastatic reproduction of the original notes for the
lectures delivered by Majorana at the state University of Neaples
(during the first months of 1938) is in:

\begin{enumerate}
\item[{[17]}] {\em Ettore Majorana -- Lezioni all'Universit\`a di
Napoli}, edited by B. Preziosi (Bibliopolis; Neaples, 1987);
\end{enumerate}

\noi while the complete set of the lecture notes (including the
so-called Moreno Document) has been published in:

\begin{enumerate}
\item[{[18]}] {\em Ettore Majorana -- Lezioni di Fisica Teorica},
edited by S. Esposito (Bibliopolis; Neaples, 2006).
\end{enumerate}

\noi See also:
\begin{enumerate}
\item[{[19]}] A. Drago and S. Esposito, ``Ettore Majorana's course
on theoretical physics: A recent discovery,'' in {\em Phys.
Persp.}, in press [e-print physics/0503084].
\end{enumerate}

\noi An English translation of (only) his notes for his Inaugural
Lecture appeared as:

\begin{enumerate}
\item[{[20]}] B. Preziosi and E. Recami, ``The preliminary notes
by E.Majorana for his Inaugural Lecture", in Ref. {[13]},
pp.263-282: e-print arXive:0709.0941[physics.hist-ph].
\end{enumerate}

\noi Other previously unknown scientific manuscripts by Majorana
have been revaluated (and/or published with comments) in the
following articles:

\begin{enumerate}
\item[{[21]}] R. Mignani, M. Baldo and E. Recami, ``About a
Dirac--like equation for the photon, according to Ettore
Majorana,'' {\em Lett. Nuovo Cim.} {\bf 11} (1974) 568; E.
Giannetto, ``A Majorana-Oppenheimer formulation of Quantum
Electrodynamics,'' {\em Lett. Nuovo Cimento}  {\bf 44} (1985) 140
and 145; \ ``Su alcuni manoscritti inediti di E. Majorana,'' in
{\em Atti del IX Congresso Nazionale di Storia della Fisica},
edited by F. Bevilacqua (Milan, 1988), p.173;  S. Esposito,
``Covariant Majorana formulation of Electrodynamics,'' {\em Found.
Phys.} {\bf 28} (1998) 231;

\item[{[22]}] S. Esposito, ``Majorana solution of the Thomas-Fermi
equation,'' {\em Am. J. Phys.} {\bf 70} (2002) 852; \ ``Majorana
transformation for differential equations,'' {\em Int. J. Theor.
Phys.} {\bf 41} (2002) 2417;  \ ``Fermi, Majorana and the
statistical model of atoms,'' {\em Found. Phys.} {\bf 34} (2004)
1431;

\item[{[23]}] E. Majorana, ``Ricerca di un'espressione generale
delle correzioni di Rydberg, valevole per atomi neutri o ionizzati
positivamente'' (Search of a general expression for Rydberg's
correction, valid for neutral or positively charges atoms), {\em
Nuovo Cimento} {\bf 6} (1929) 14-16. \ The corresponding original
material is all contained in Refs.{[15,16]}, while a comment is
in S. Esposito, ``Again on Majorana and the Thomas-Fermi model: A
comment about physics/0511222,'' [arXiv:physics/0512259];

\item[{[24]}] S. Esposito, ``A peculiar lecture by Ettore
Majorana,'' {\em Eur. J. Phys.} {\bf 27} (2006) 1147; \ ``Majorana
and the path-integral approach to Quantum Mechanics,'' {\em Ann.
Fond. Louis de Broglie} {\bf 31} (2006) 1;

\item[{[25]}] S. Esposito, ``Hole theory and Quantum
Electrodynamics in an unknown manuscript in French by Ettore
Majorana,'' {\em Found. Phys.}, in press [arXiv:physics/0609137].
\end{enumerate}

\noi Some scientific papers, elaborating on several intuitions by
Majorana, are for instance the following:

\begin{enumerate}
\item[{[26]}] D. Fradkin, ``Comments on a paper by Majorana
concerning elementary particles,'' in {\em Am. J. Phys.} {\bf 34}
(1966) 314;

\item[{[27]}] R. Penrose, ``Newton, quantum theory and reality,''
in {\em 300 Years of Gravitation}, edited by S.W. Hawking and W.
Israel (Cambridge University Press; Cambridge, 1987); \ J. Zimba
and R. Penrose, {\em Stud. Hist. Phil. Sci.} {\bf 24} (1993) 697;
R. Penrose, {\em Ombre della Mente (Shadows of the Mind)}
(Rizzoli; Milan, 1996), pp.338--343 and 371--375;

\item[{[28]}] C. Leonardi, F. Lillo, A. Vaglica and G. Vetri,
``Majorana and Fano alternatives to the Hilbert space,'' in {\em
Mysteries, Puzzles, and Paradoxes in Quantum Mechanics}, edited by
R. Bonifacio (A.I.P.; Woodbury, N.Y., 1999) 312; \ `` Quantum
visibility, phase-difference operators, and the Majorana Sphere,''
preprint (Phys. Dept., Univ. of Palermo, Italy; 1998); \ F. Lillo,
``Aspetti fondamentali nell'in\-ter\-fe\-ro\-me\-tria a uno e due
Fotoni,'' Ph.D. thesis (C. Leonardi supervisor), Dept. of Physics,
University of Palermo, 1998;

\item[{[29]}] R. Casalbuoni, ``Majorana and the infinite component
wave equations,'' [arXiv:hep-th/0610252].
\end{enumerate}

\noi Further scientific papers can be found in:

\begin{enumerate}
\item[{[30]}] I. Licata (editor), ``Majorana Legacy in
Contemporary Physics'', special issue no.10 of the {\em Electr. J.
Theor. Phys.} {\bf 3} (2006); V. Dvoeglazov (editor), special
issue nos.2-3 of the {\em Annales de la Fondation Louis de
Broglie} {\bf 31} (2006).
\end{enumerate}

\noi Further historical studies on Majorana's work may be found in
the following recent papers:

\begin{enumerate}
\item[{[31]}] A. De Gregorio, ``Il `protone neutro', ovvero della
laboriosa esclusione degli elettroni dal nucleo,'' in {\em Physis}, in press
[arXiv:physics/ 0603261];

\item[{[32]}] A. De Gregorio and S. Esposito, ``Teaching
Theoretical Physics: The cases of Enrico Fermi and Ettore
Majorana,'' in {\em Am. J. Phys.}, in press,
[arXiv:physics/0602146];

\item[{[33]}] E. Di Grezia and S. Esposito, ``Majorana and the
quasi-stationary states in nuclear physics''
[arXiv:physics/0702179];

\item[{[34]}] A. Drago and S. Esposito, ``Following Weyl on
quantum mechanics: The contribution of Ettore Majorana,'' {\em
Found. Phys.} {\bf 34} (2004) 871.
\end{enumerate}

\noi Last but not least, the English translation of a selection of the {\em Quaderni}
is going to be published in:

\begin{enumerate}
\item[{[35]}] S.Esposito, E.Recami, A.van der Merwe, and R.Battiston (editors), {\em Ettore
Majorana -- Unpublished Research Notes on Theoretical Physics} (Springer;
Berlin, to appear).
\end{enumerate}

\end{document}